# High-order Harmonic Generation and its Unconventional Scaling Law in the Mott-insulating $Ca_2RuO_4$

K. Uchida[1,*], G. Mattoni[1], S. Yonezawa[1], F. Nakamura[2], Y. Maeno[1], K. Tanaka[1,*]

[1]*Department of Physics, Graduate School of Science, Kyoto University, Kyoto, Kyoto 606-8502, Japan*

[2]*Department of Education and Creation Engineering, Kurume Institute of Technology, Kurume, Fukuoka 830-0052, Japan*

*Correspondence should be addressed K. U. (uchida.kento.4z@kyoto-u.ac.jp) or K. T. (kochan@scphys.kyoto-u.ac.jp ).

## Abstract

Competition and cooperation among orders is at the heart of many body physics in strongly correlated materials and leads to their rich physical properties. It is crucial to investigate what impact many-body physics has on extreme nonlinear optical phenomena, with the possibility of controlling material properties by light. However, the effect of competing orders and electron-electron correlations on highly nonlinear optical phenomena has not yet been experimentally clarified. Here, we investigated high-order harmonic generation from the Mott-insulating phase of $Ca_2RuO_4$. Changing the gap energy in $Ca_2RuO_4$ as a function of temperature, we observed a strong enhancement of high order harmonic generation at 50 K, increasing up to several hundred times compared to room temperature. We discovered that this enhancement can be well-reproduced by an empirical scaling law that depends only on the material gap energy and photon emission energy. Such scaling law cannot be explained by a simple two-band model under the single electron approximation. Our results suggest that the highly nonlinear optical response of strongly correlated materials is deeply coupled to their electron-electron correlations and resultant many-body electronic structure.

Strong coupling among multiple degrees of freedom often results in a variety of electronic ground states, especially in strongly correlated materials. A strong light field may drive such systems far from equilibrium and modify the balance among competing orders. After light irradiation, the material can relax to a non-trivial meta-stable state such as a transient superconducting state or a ferroelectric state [1-4]. Under a strong light field, coherent and highly nonlinear optical phenomena can also be observed. High order harmonic generation (HHG) is the most typical of such phenomena [5-15], where coherent photons with integer multiples of the driving photon energy are emitted. HHG has provided fundamental insights into non-perturbative light-matter interactions in solids [13]. So far, HHG has mainly been studied in systems that can be described with the single electron approximation, giving information about the band structure [7,8], inter-atomic bonding [9,14], Berry curvature [11], and transition dipole moment [15]. HHG spectroscopy is applicable to strongly correlated systems, where electron-electron correlations play a dominant role, and has the potential to be a new method to elucidate the dynamical aspects of many-body physics. Several theoretical studies predict the characteristics of HHG for strongly correlated materials [16-19]. However, only a few HHG experiments have been performed so far on strongly correlated materials [20,21], and the unique properties related to their electronic correlations has yet to be found.

In this Letter, we report HHG from the Mott-insulating phase of $Ca_2RuO_4$ with a strong enhancement of HHG yields upon lowering temperature. The strongly temperature-dependent gap energy of $Ca_2RuO_4$ allows us to investigate the relationship between HHG emission and the gap energy in this strongly correlated system. We find that the yields of the high harmonics scale with the gap energy and the high harmonic emission energy. The observed scaling law cannot be reproduced by a simple two-band model under the single electron approximation. This suggests that electron-electron correlations and the resultant complex electronic structure in $Ca_2RuO_4$ play a crucial role in highly nonlinear optical response.

$Ca_2RuO_4$ is a layered perovskite which shows a variety of electronic properties owing to the competition between multiple degrees of freedom [22,23]. The material is a Mott insulator below the metal-insulator transition temperature $T_{MI}$ ~360 K and shows antiferromagnetic order below $T_N$ ~110 K [24]. The electronic properties can be controlled by varying external parameters such as chemical substitution [25], pressure [26], electric field [27], and temperature [24].

The gap energy of $Ca_2RuO_4$ changes from 0.2 eV to 0.65 eV upon lowering temperature below $T_{MI}$ [28].

We used an intense mid-infrared (MIR) pulse (central photon energy: $\hbar\Omega_{MIR} = 0.26$ eV, repetition rate: 1 kHz, and pulse duration: 100 fs) to measure HHG emissions from samples of $Ca_2RuO_4$. Figure 1(a) shows a schematic view of the experimental setup [29]. The sample was set in an optical cryostat, and its temperature was varied between 290 K and 50 K within $Ca_2RuO_4$ Mott insulating phase. MIR pulses were focused onto the sample using a reflective-type objective lens and passed through a 1-mm-thick $CaF_2$ window. The typical MIR intensity at the focus point in vacuum is 0.3 TW/cm², corresponding to an electric field strength of 15 MV/cm, and pulse energy of around 0.3 μJ (spot diameter 27 μm). Since bulk $Ca_2RuO_4$ is not transparent above the gap energy, we collected the HHG signal in reflection geometry using the same objective lens. In this configuration, we have the advantage of a negligible nonlinear propagation effect that in other configurations would strongly distort the spectral shape [30-32]. Figure 1(b) shows an optical image of a single crystal $Ca_2RuO_4$ sample with exposed ab-plane. The transition temperatures ($T_{MI}$=360 K and $T_N$=110 K) were confirmed by magnetic susceptibility measurements [29]. The crystal orientation was determined by X-ray diffraction as well as by magnetic anisotropy measurements. Throughout the experiment, we used a linearly polarized MIR pulse with polarization lying in the ab plane, and electric field direction (polarization) was identified by the angle θ with respect to the b axis (Fig. 1(b)).

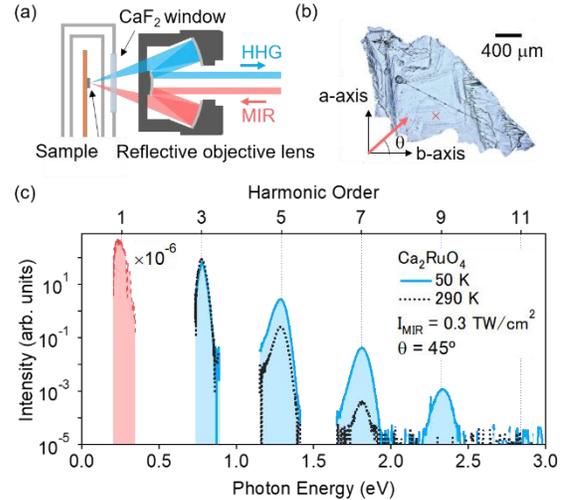

FIG 1 High harmonic generation in $Ca_2RuO_4$. (a) Schematic diagram of the experimental setup for HHG measurement in reflection geometry (extended details in Fig. S3). (b) Optical image of the $Ca_2RuO_4$ sample. The position of the MIR spot, that is focused on a region where the sample surface roughness is small, is indicated by the red cross mark (spot diameter 27 μm). The red arrow indicates the direction of the MIR polarization which is identified by the angle θ that it forms with the b-axis. (c) Typical HHG spectra obtained at 50 K (light-blue shaded area) and 290 K (black dotted line) with MIR intensity of 0.3 TW/cm² and polarization θ = 45°. The red shaded area indicates the incident MIR spectrum, measured using a monochromator equipped with a HgCdTe detector, that is decreased by a factor of $10^{-6}$ in the plot for clarity.

Figure 1(c) shows typical HHG spectra obtained at 290 K (black dotted line) and 50 K (light blue line and shaded area) with an incident MIR intensity $I_{MIR}$ of 0.3 TW/cm$^2$ and $\theta = 45°$. We observed only odd-order harmonics because of the inversion symmetry of Ca$_2$RuO$_4$. While the third-order harmonic yield is just slightly suppressed at low temperature, we observe a strong enhancement of the fifth and higher harmonics. The enhancement is stronger for larger harmonic order, reaching several hundred times yield for the ninth harmonic at 50K compared to 290 K.

Changes of Ca$_2$RuO$_4$ optical properties in the MIR region can change the excitation conditions, possibly accounting for part of this enhancement. However, we estimated that the change of the MIR electric field inside the sample due to its reflectivity change between 290 K and 50 K is less than 7 % [29]. As it will be shown in Figs. 2(a-c), this change can only explain an increase of the HHG signal of just about 25 %. This fact indicates that the observed enhancement in HHG intensity has a different origin, possibly intrinsic to the electronic structure of the material. As another possibility, HHG yields is enhanced by out-going resonances, which has been reported in semiconductor [33], indicating a strong connection between the HHG spectrum and optical conductivity. However, even if the linear optical properties of Ca$_2$RuO$_4$ change with temperature [28, 34], only small changes occur above 1.3 eV, where we observed the strong HHG enhancement. Hence, the strong HHG enhancement indicates that the nonlinearity of the system is increased at low temperature.

We now study the dependence of the $n$th-order harmonic intensity $I_n$ on the incident MIR intensity $I_{MIR}$ as shown in Figs. 2(a)-(c). There are strong deviations from the predictions of perturbative nonlinear optics ($I_n \propto I_{MIR}^n$, dotted lines), because all HHG intensities are almost proportional to the square of MIR intensity ($I_n \propto I_{MIR}^2$). This indicates that non-perturbative light-matter interactions take place under our experimental conditions. Moreover, the dependences of the normalized HHG intensities on the MIR intensity are almost unchanged in the studied temperature range. These results suggest that the microscopic HHG mechanism is unaffected by changes in temperature.

Figures 2(d)-(f) show the crystal-orientation dependence of the third, fifth, and seventh harmonic intensities at 290 K and 50 K. All the harmonics show maximum values for θ~45°, for which the MIR electric field points along the direction from Ru to nearest-neighbor Ru or O atoms (Fig. 2(g)). This suggests that inter-atomic transitions such as inter-site d-d transitions or charge transfer transitions may be involved in the HHG process [9,17]. Since HHG emission energies are

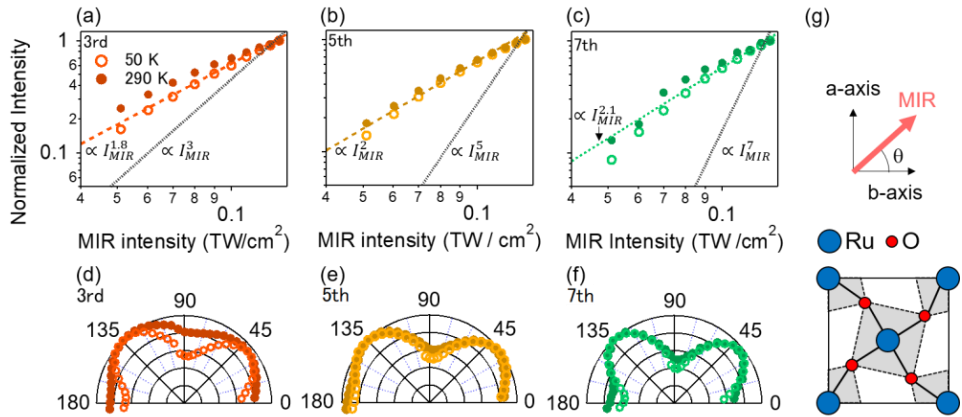

FIG 2 Dependence of the HHG yields on MIR intensity and polarization. (a-c) Normalized (a) third, (b) fifth, and (c) seventh harmonic intensity as a function of MIR intensity with MIR polarization angle $\theta = 45°$. Open circles measured at 50 K, solid cicles measured at 290 K. All the harmonic intensities are normalized by their peak value. Dashed lines are guides to the eye which are proportional to $I_{MIR}^{1.8}$, $I_{MIR}^2$, $I_{MIR}^{2.1}$. The black dotted lines indicate the trends expected from perturbation theory ($I_n \propto I_{MIR}^n$, $n$: harmonic order). (d-f) Polar plots of the (d) third, (e) fifth, and (f) seventh harmonic intensity as a function of MIR polarization angle θ. The MIR intensity was fixed at the maximum value of 0.3 TW/cm$^2$. (g) Schematic of the Ru-O plane in Ca$_2$RuO$_4$ respect to which the polarization is changed. The grey shaded areas indicate the oxygen octahedra.

in the region of the d-d transition [28], the anisotropy of HHG yields may be attributed to d-d transitions between nearest-neighbor Ru ions. By lowering the temperature, the HHG anisotropy becomes stronger. However, this increased anisotropy is not salient in the fifth and seventh harmonics, whereas strong enhancements in their HHG yields are observed. These results support the conclusion that the HHG mechanism is not directly caused by changes in temperature.

To investigate the temperature dependence in more detail, we measured HHG intensities with $\theta = 45°$ as a function of temperature. Normalized intensities of $n$th-order harmonics $I_n(T)$ are shown in Figs. 3(a)-(d) (solid circles). HHG intensities in $Ca_2RuO_4$ do not show any noticeable variation in proximity of the Neel transition ($T_N$ = 110 K), indicating that magnetic order does not dominantly affect the HHG mechanism. The third harmonics slightly decreases by lowering temperature. For the fifth and higher harmonics, instead, HHG intensities exponentially increase down to about 200 K and then show a gradual saturation. The rate of increase of HHG signals is larger for higher harmonics. For comparison, we also measured HHG from the conventional narrow-gap semiconductor InAs (open squares in Figs. 3), which has a gap energy (~0.3 eV) and carrier concentration (~ $10^{16}$ cm$^{-3}$) similar to those of $Ca_2RuO_4$ at room temperature [35]. In contrast to $Ca_2RuO_4$, the HHG intensities that we measured for InAs are almost independent of temperature. This implies that the HHG mechanism in Mott-insulating $Ca_2RuO_4$ is qualitatively different from that of semiconductors or band insulators.

To investigate the possible effect of carrier concentration in the HHG process, we measured the temperature dependence of HHG yields in La-doped $Ca_2RuO_4$ ($Ca_{2-x}La_xRuO_4$, x=0.016). The substitution of $Ca^{2+}$ for $La^{3+}$ causes an increase in carrier concentration of about four orders of magnitude at room temperature and a decrease of $T_{MI}$ [36]. Despite these substantial changes, strong enhancement of HHG yields at low temperature remains qualitatively unchanged, and HHG spectrum in La-doped sample can be understood as the reduction of gap energy by chemical substitution [29]. This indicates that the effect of activated carriers on HHG yields is negligible, while a strong link is expected with the change in gap energy.

We found that the observed enhancement in HHG yields at low temperature has a direct correlation with the optical gap energy $2\Delta$ of $Ca_2RuO_4$. By using the temperature dependence $2\Delta(T)$ found in literature [28], we plot HHG intensities as a function of gap energy in

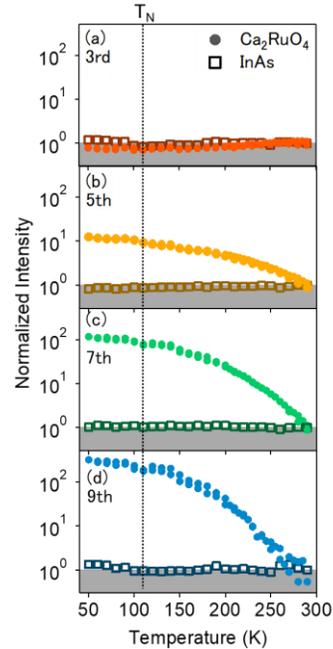

FIG 3 Enhancement of HHG at low temperature. (a-d) Normalized (a) third, (b) fifth, (c) seventh, and (d) ninth harmonic intensity as a function of sample temperature with MIR intensity of 0.3 TW/cm$^2$ at $\theta = 45°$. Solid circles (open squares) indicate the HHG observed in $Ca_2RuO_4$ (InAs). Dashed lines at 110 K indicate the Neel temperature $T_N$.

Fig. 4(a). For the third-order harmonics, the intensity slightly decreases with increasing gap energy. The fifth and higher harmonic intensities, instead, increase with the gap energy. Moreover, HHG yields can be well described as exponential functions of $2\Delta(T)$. This clearly indicates a strong connection between the gap energy and HHG enhancement in $Ca_2RuO_4$. Surprisingly, our experimental results are contrary to what expected from the carrier tunneling process, according to which an increase in gap energy suppresses carriers generated by the field tunneling process, diminishing the resulting HHG signal [36-39]. This suggests that increasing the gap energy enhances the nonlinearity of the system to an extent that overcomes the attenuation of tunneling carriers.

To obtain more information about the enhancement in HHG yields, we measured HHG with different incident MIR photon energies [29]. Figure 4(b) shows the ratio of HHG intensity at 290 K to that at 50 K, which we define the enhancement factor, as a function of HHG emission photon energy $\hbar\omega_e$. We compare the results obtained with two different incident MIR photon energies: $\hbar\Omega_{MIR} = 0.26$ eV (brown stars) and 0.19 eV (purple crosses). Within the same harmonic order, the enhancement factor decreases as the incident photon energy decreases. Surprisingly, the enhancement factor follows a single power law ($\propto (\hbar\omega_e)^l, l \sim 5.6$) regardless of the incident photon energy. This indicates that the enhancement in HHG yields is not determined by the driving photon energy, but rather by the emission photon energy.

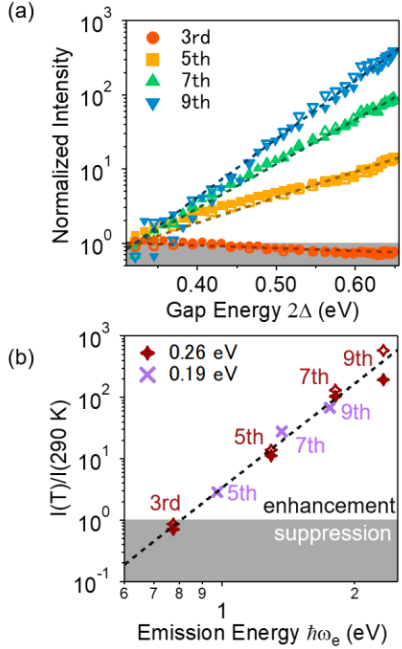

FIG 4 Empirical scaling law of HHG enhancement in $Ca_2RuO_4$. (a) Normalized HHG intensity as a function of optical gap energy of $Ca_2RuO_4$ $2\Delta$ from literature [28]. Orange circles, yellow squares, green triangles, and blue inverted triangles indicate the third, fifth, seventh, and ninth harmonics, respectively. Open (solid) polygons were measured with increasing (lowering) sample temperature. The dashed lines indicate the fitting curves by using Eq. (1) with $\hbar\Omega_{th} = 0.8$ eV and $2\Delta_{th} = 58$ meV. All the harmonic intensities are normalized by the values at 290 K. (b) Enhancement factor defined as the ratio of HHG intensity at 50 K to 290 K $I_n(50 K)/I_n(290 K)$ as a function of emission energy. Brown stars and purple crosses are measured with an MIR photon energy of 0.26 eV and 0.19 eV, respectively. Solid (open) symbols were measured with increasing (lowering) temperature. The dashed line indicates the fitting curve given by Eq. (1).

We propose an empirical scaling law that satisfies the characteristics of both the two experimental result in Figs. 4(a) and (b), where the *n*th order harmonic yield as a function of gap energy $2\Delta(T)$ and emission energy $\hbar\omega_e (= n\hbar\Omega_{MIR})$ can be expressed by

$$I_n(T)/I_n(T_0) \approx \left(\frac{n\Omega_{MIR}}{\Omega_{th}}\right)^{\frac{\Delta(T)-\Delta(T_0)}{\Delta_{th}}}. \qquad (1)$$

Here, $\hbar\Omega_{th}$ and $2\Delta_{th}$ are threshold values of the HHG emission energy and gap energy, respectively. We use Eq.(1) to fit the data in both Figs.4(a) and (b) (dashed lines) and find good agreement with the experiment by using the only two parameters $\hbar\Omega_{th} \sim 0.8$ eV and $2\Delta_{th} \sim 58$ meV. This empirical equation allows to calculate the harmonic yields at temperature T when the HHG yields at a fixed temperature $T_0$ are known. Remarkably, Eq. (1) describes both the slight suppression of the third harmonic and the enhancement of the fifth and higher harmonics. The fact that it is not the harmonic order but rather the HHG emission energy that determines the enhancement factor implies that the observed enhancements are regulated by the emission process. This simple, but rather extraordinary scaling law of HHG yields has never been reported in previous studies on HHG in solids.

We performed a numerical calculation confirming that changes in the bandgap energy in a conventional band insulator do not lead to such behavior in HHG yields. In contrast to the observed HHG yields, the calculated HHG signals show non-monotonic behavior as a function of the gap energy, and do not show any exponential enhancement (Figs. S8 and S9) [29], which was also confirmed experimentally on InAs (Fig. S10) [29]. This implies that electron-electron correlation or the resultant modification of the electronic structure, which are neglected in our simulation, play an important role in enhancing HHG yields in $Ca_2RuO_4$.

Previous theoretical studies of HHG considering the electron correlation effect have reported that the total high-harmonic emission spectrum shifts toward higher photon energy as the gap energy increases [18,19]. Also, the analytical formulation of the nonlinear optical response in strongly correlated systems predicts that higher-order response is strongly enhanced by the correlation effect in a perturbative manner [40]. These findings are qualitatively similar to our experimental results, i.e., stronger enhancement of higher harmonics upon increasing the material gap energy. However, in our experiment, the enhancement is more likely to be determined not by the harmonic order, but by the emission energy. To understand the observed scaling law, more refined theoretical studies on the effect of the gap energy on HHG are needed.

In conclusion, we investigated the properties of HHG emission from Mott-insulating $Ca_2RuO_4$. HHG efficiencies whose orders are greater than the third show strong enhancement as the temperature is lowered. We found that the observed temperature dependence is strongly linked to temperature-driven changes of $Ca_2RuO_4$ gap energy, with an exponential enhancement in HHG yields as the gap energy increases. This enhancement does not occur in ordinary semiconductors, suggesting that electron-electron correlations play an important role in the HHG mechanism in $Ca_2RuO_4$. The HHG enhancements observed in this work are well described by the empirical scaling law expressed by Eq. (1) that depends only on the material gap energy and HHG emission energy. This scaling law suggests that the HHG process in $Ca_2RuO_4$ underlies a physics peculiar to strongly correlated systems. Further theoretical and experimental studies, focusing specifically on the nonlinear optical response of strongly correlated materials, may help to elucidate the possible universality of the observed scaling law. These undertakings will lead to a better understanding of strongly correlated materials, possibly enabling us to control many-body structures on an ultrafast time scale and ultimately to design novel functional materials.

**Acknowledgements**
The authors are thankful to Hideki Narita for characterizing La-doped $Ca_2RuO_4$ samples. K.U. and K.T. are thankful to Yuta Murakami and Kazuaki Takasan for fruitful discussions. This work was supported by Grant-in-Aid for Scientific Research (S) (Grant Nos. JP17H06124 and JP17H06136), JST ACCEL Grant (No. JPMJMI17F2), and the JSPS Core-to-Core Program (No. JPJSCCA20170002). K. U. is thankful for a Grant-in-Aid for Young Scientists (Grant No. 19K14632). G. M. acknowledges the support from the Dutch Research Council (NWO) through Rubicon Grant No. 019.183EN.031.

# Supplemental Information:

# High-order Harmonic Generation and its Unconventional Scaling Law in the Mott-insulating $Ca_2RuO_4$

K. Uchida[1,*], G. Mattoni[1], S. Yonezawa[1], F. Nakamura[2], Y. Maeno[1], K. Tanaka[1,*]

[1]*Department of Physics, Graduate School of Science, Kyoto University, Kyoto, Kyoto 606-8502, Japan*

[2]*Department of Education and Creation Engineering, Kurume Institute of Technology, Kurume, Fukuoka 830-0052, Japan*

## 1. Sample preparation and characterization.

Single crystals of $Ca_2RuO_4$ were grown by the floating-zone method with $RuO_2$ self-flux using a commercial infrared furnace (Canon Machinery, model SC-M15HD). We used a high-density feed rod prepared by calcinating a mixture of $CaCO_3$ (99.99%) and $RuO_2$ (99.9%) in air at 1000 °C for 10 hours. The crystals were grown at the rate of ~10 mm/h in a gas mixture of 10% O2 and 90% Ar at a pressure of 1 MPa. X-ray diffraction measurements indicate that the crystals are of single phase of $K_2NiF_4$ structure (space group S-Pbca) with the lattice parameter c of 11.915 Å at room temperature. We note that $Ca_2RuO_4$ crystals that we prepared under the same conditions exhibit a metallic state with a residual resistivity of only ~3 μΩ cm with pressure above 2 GPa [1]. This value is very

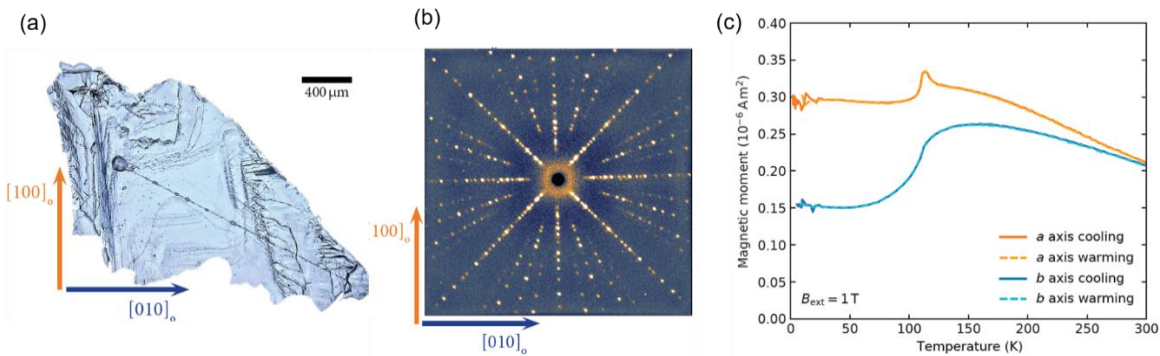

Fig. S1 (a) Optical image of the sample used in our experiment. (b) Laue diffraction pattern of the sample a-b surface used for the experiment. (c) Magnetization as a function of temperature measured with a constant magnetic field of 1T upon cooling and warming.

low for the resistivity of an oxide metal, indicating that our samples are of high quality with low density of defects. The single crystals of $Ca_2RuO_4$ used in this work are high-purity with essentially stoichiometric oxygen content.

Figure S1(a) shows the optical image of the high-quality single crystalline sample (Sample CR53_2_G2) used in our experiment. The crystal orientation of the sample was confirmed by x-ray diffraction measurement as shown in Fig. S1(b). Field-cooled (-heated) magnetization measurements with $B_{ext}$ along a-axis ($[100]_o$) show a peak at 113 K, indicating the insurgence of antiferromagnetic order. This temperature is in good agreement with the Neel temperature of S-phase $Ca_2RuO_4$.

## 2. Temperature dependence of reflectivity at incident photon energy.

In order to estimate the mid-infrared (MIR) electric field inside the sample, we measured reflectivity spectra by changing the sample temperature. Figure S2(a) shows reflectivity spectra obtained by using Fourier transform infrared spectrometer at 285 K (orange dashed line), 205 K (green dotted line), and 60 K (blue solid line). The spectral shape and their absolute value are in good agreement with previous reports and show gradual decrease upon lowering temperature [2]. Figure S2(b) shows the reflectivity at 0.26 eV as a function of sample temperature. The reflectivity varies from 0.26 to 0.16 by decreasing the temperature from 295 K to 60 K. Since the absorption in this frequency range is negligible [2], the ratio of electric field inside the sample at 285 K to 60 K can be estimated to be ~93%. In our experimental conditions, HHG intensities are nearly proportional to the square of incident MIR intensity (see Figs. 2 in the main text). Therefore, the temperature dependence of the reflectivity gives only a small contribution to the strong enhancement in HHG yields as a function of temperature.

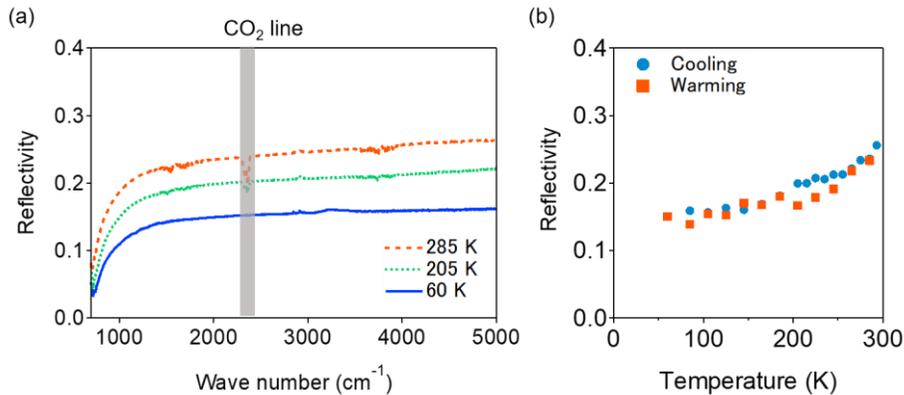

Fig. S2 (a) Reflectivity spectra at 285 K (orange dashed line), 205 K (green dotted line), and 60 K (blue solid line) taken by FTIR spectrometer. (b) Reflectivity at 0.26 eV as a function of sample temperature. Light-blue circles (orange squares) were taken upon cooling (warming).

## 3. Experimental setup of HHG measurement.

Figure S3 shows the schematic of our experimental setup. We used a Ti: Sapphire regenerative amplifier (pulse width: 35 fs, pulse energy: 7 mJ, center wavelength: 800nm, repetition rate: 1 kHz) as laser source. The laser output (~4 mJ) is used for generating the MIR driving field. We first generated the signal outputs with center wavelengths of 1180 nm and 1570 nm by using a dual optical parametric amplifier system (Light Conversion TOPAS-TWINS), and then generated MIR pulses (center wavelength: 4.8 μm) by difference frequency mixing the signals in a GaSe crystal. The transmitted input signals are blocked by using a longpass filter (cutoff wavelength: 4 μm). Polarization angle and intensity of the MIR pulses are controlled by three wire-grid polarizers and a liquid crystal variable retarder (Thorlabs LCC1113-MIR). MIR pulses were reflected by an indium tin oxide (ITO) plate, which shows high reflectivity in the MIR region and high transmissivity in the visible region. Then, MIR pulses were focused onto the sample by using a reflective-type objective lens, with effective focal length of 13 mm (working distance 24 mm). The spot size at the focal point was estimated to be 27 μm in full-width at half-maximum by using a knife edge measurement, while the pulse width is estimated to be 100 fs by using electro-optic sampling. The samples were set into a cryostat with a 1 mm-thick $CaF_2$ window, and the sample position was controlled by mechanical stages so that it located at the center of MIR spot. The reflected high harmonic emissions were collected using the same objective lens and passed through the ITO plate. The emissions were spectrally resolved by a spectrometer, and detected by an InGaAs line detector (for the third harmonics) or a Si CCD camera (for the higher harmonics).

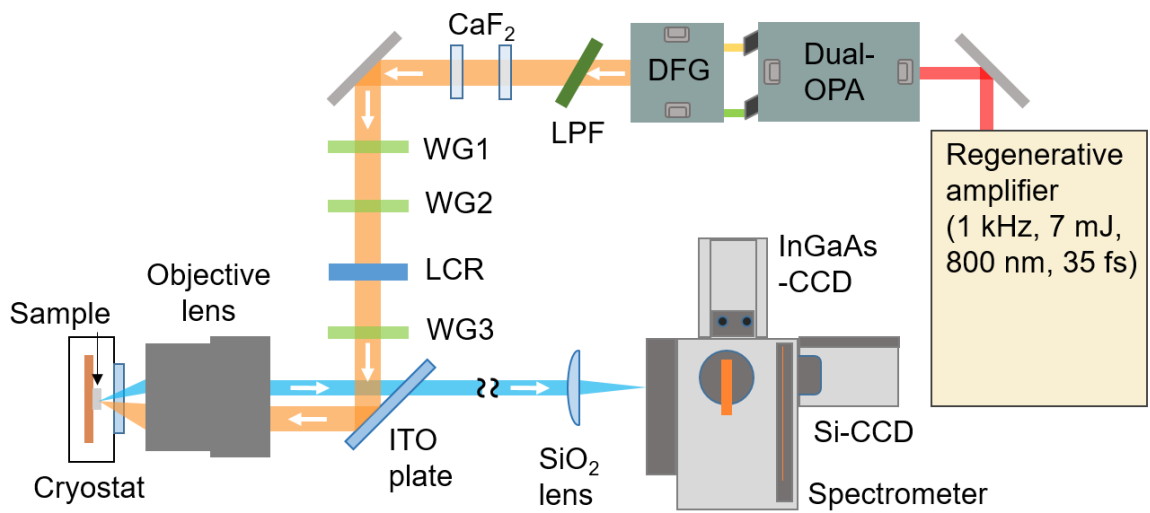

Fig. S3 Schematics of the HHG experimental setup. A pair of CaF$_2$ plates is used to compensate the group velocity dispersion of all the optics that the MIR pulses pass through. LPF: lowpass filter. WG: wiregrid polarizer. LCR: liquid crystal retarder.

## 4. La-doping effect.

To investigate the effect of slight chemical substitution on the HHG properties, we observed HHG from La-doped $Ca_2RuO_4$. The slight doping of La causes the relaxation of the distortions of the $RuO_6$ octahedra and results in a decreased metal-insulator transition temperature $T_{MI}$, with a concomitant reduction of the optical gap[3]. Here, we used $Ca_{2-x}La_xRuO_4$ with x=0.016 and $T_{MI}$ = 325 K, measured using energy dispersive X-ray spectroscopy and magnetic susceptibility, respectively. In addition to the suppression of $T_{MI}$, Figures S4(a) and (b) respectively show the HHG spectra from the pristine $Ca_2RuO_4$ and La-doped $Ca_2RuO_4$. Enhancement of the higher harmonics with lowering temperature is observed in both cases. Furthermore, we note that in the La-doped case the HHG spectrum is similar to the pristine case, but shifted to lower temperatures. This behavior can be understood as due to the reduction of gap energy by chemical substitution, causing the reduction of HHG yields for higher harmonics. Figures S5 show the normalized intensities of (a) the third, (b) fifth, and (c) seventh harmonics as a function of the temperature. HHG yields for the La-doped sample show similar behavior to those for the pristine sample. This indicates that our observed relationship between HHG yields and gap energy is robust against chemical substitution.

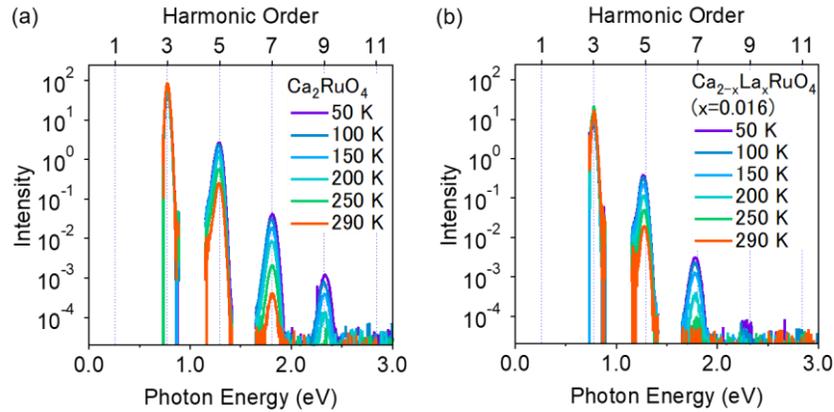

Fig. S4 (a) HHG spectra from pristine $Ca_2RuO_4$ at several temperatures (50 K, 100 K, 150 K, 200 K, 250 K, 290 K. (b) HHG spectra from La-doped $Ca_2RuO_4$ at the same temperatures.

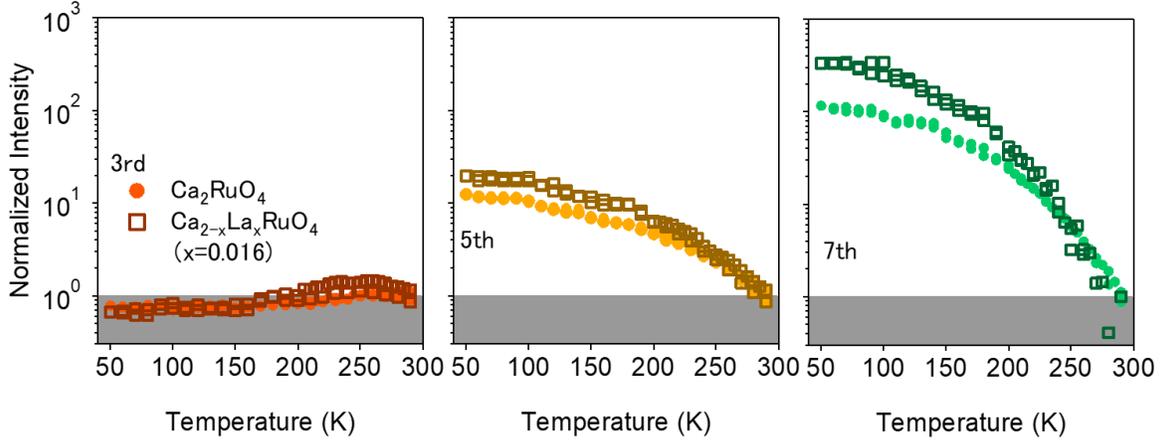

Fig. S5 (a) third, (b) fifth, and (c) seventh harmonic intensities of pure $Ca_2RuO_4$ (solid circles) and La-doped $Ca_2RuO_4$ (open squares) as a function of temperature. Here, the MIR photon energy is set to 0.26 eV. HHG intensities are normalized by the values at 290 K.

5. **Incident MIR photon energy dependence.**

We investigated the effect of driving MIR photon energy on HHG yields. Figures S6 (a) and (b) show HHG spectra at 50 K with MIR photon energies of 0.26 eV and 0.19 eV, respectively. HHG intensities gradually decrease with an increase of harmonic order for both 0.26 eV and 0.19 eV excitations, and approximately follow the exponential curves (dashed lines). These simple HHG spectra indicate that out-going resonant enhancement of HHG yields owing to a certain electronic transition, which is observed in some band insulator [4, 5], is not salient in $Ca_2RuO_4$.

One possible factor that can influence the HHG yields is the ratio of the gap energy to the driving MIR photon energy ($2\Delta/\hbar\Omega_{MIR}$). This is because the electronic transition (or tunneling) between states separated by the gap, which is the first step occurring in HHG processes, should depend on $2\Delta/\hbar\Omega_{MIR}$. Especially, for sub-gap excitation condition ($2\Delta/\hbar\Omega_{MIR} > 1$), a decrease of incident photon energy suppress the transition to the excited state. However, as shown in Figs. S6 (a) and (b), MIR photon energy dependence of HHG yields at 50 K is not salient within the same harmonic order, implying that carrier generation above the gap is not a crucial factor for HHG yields in our experimental setup.

Figure S7 shows HHG intensities as a function of temperature with MIR photon energy of 0.19 eV. The observed behavior is almost the same as with 0.26 eV (Fig. 3 in the main text), but the amplitude of the enhancement is relatively weaker within the same harmonic order. As shown in Fig. 4(b), this reflects that the enhancement of HHG yields is scaled by emission energy of the high harmonics.

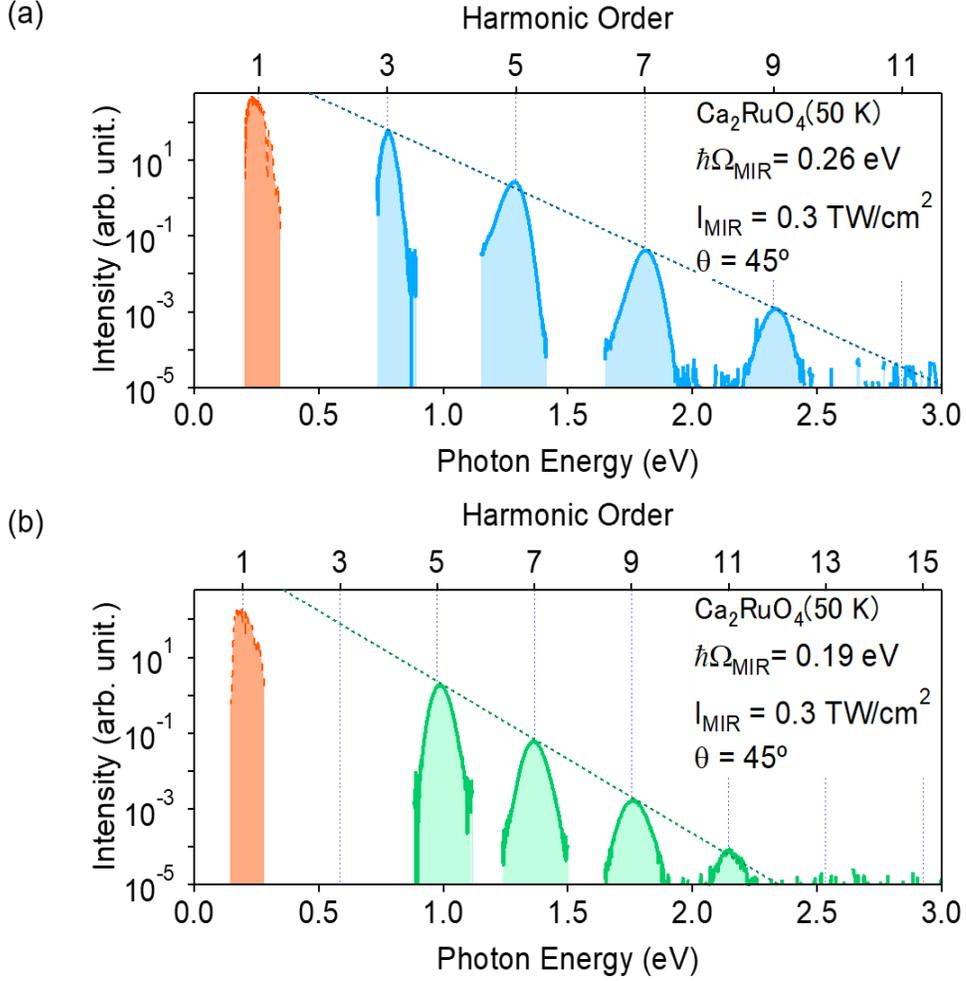

Fig. S6 (a) HHG spectrum at 50 K with incident photon energy of 0.26 eV (blue shaded area). The dotted blue line is a guide to the eye ($\propto \exp(-\omega/\omega_{th})$, $\hbar\omega_{th} = 0.14$ eV). (b) HHG spectrum at 50 K with incident photon energy of 0.19 eV (green shaded area). The dotted green line is a guide to the eye ($\propto \exp(-\omega/\omega_{th})$, $\hbar\omega_{th} = 0.11$ eV). For both (a) and (b), the MIR intensities are set at 0.3 TW/cm², and MIR polarizations are set at $\theta = 45°$. Orange shaded areas show incident MIR spectra obtained by MCT detector equipped with spectrometer, and their vertical scales are magnified for clarity. We could not observe the third order harmonics of 0.19 eV excitation due to the limitation of detectors in our experimental setup.

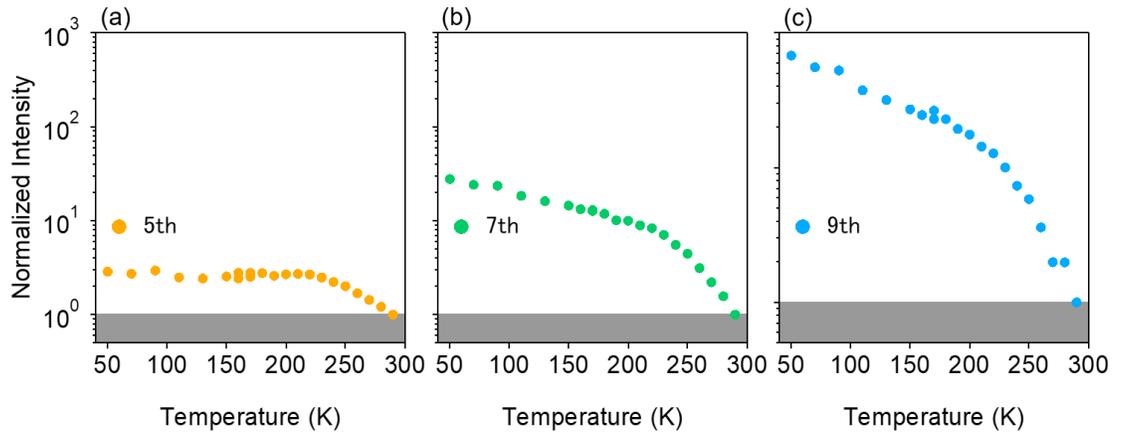

Fig. S7 (a) third, (b) fifth, and (c) seventh harmonic intensities of pristine Ca$_2$RuO$_4$ (solid circles) as a function of temperature. Here, the MIR photon energy is set at 0.19 eV. HHG intensities are normalized by the values at 290 K.

# 6. Numerical calculation of two band model.

To investigate the effect of gap energy on HHG yields, we performed the simulation of electron dynamics in a two-dimensional two-band model with cosine band structure based on Ref. [6] and Ref. [7]. This model describes HHG processes in semiconductors without electron-electron correlations.

We numerically solved the temporal evolution of the density matrix $\tilde{\rho}$ as follows:

$$\hbar \frac{d}{dt}\tilde{\rho}_{vv}(\boldsymbol{k}_0, t) = 2Im[\boldsymbol{D}_{cv}^*(\boldsymbol{K}(t)) \cdot \boldsymbol{E}(t)\tilde{\rho}_{cv}(\boldsymbol{k}_0, t)], \quad (S1)$$

$$\hbar \frac{d}{dt}\tilde{\rho}_{cc}(\boldsymbol{k}_0, t) = -\hbar \frac{d}{dt}\tilde{\rho}_{vv}(\boldsymbol{k}_0, t), \quad (S2)$$

$$\hbar \frac{d}{dt}\tilde{\rho}_{cv}(\boldsymbol{k}_0, t) = -i\big(\varepsilon_g(\boldsymbol{K}(t)) - i\gamma\big)\tilde{\rho}_{cv}(\boldsymbol{k}_0, t),$$
$$-i\boldsymbol{D}_{cv}(\boldsymbol{K}(t)) \cdot \boldsymbol{E}(t)(\tilde{\rho}_{vv}(\boldsymbol{k}_0, t) - \tilde{\rho}_{cc}(\boldsymbol{k}_0, t)), \quad (S3)$$

$$\boldsymbol{A}(t) = -\int^t dt'\, \boldsymbol{E}(t'), \quad (S4)$$

$$\boldsymbol{K}(t) = \boldsymbol{k}_0 + \frac{e}{\hbar}\boldsymbol{A}(t). \quad (S5)$$

Here, $c$ and $v$ denote the conduction and valence bands respectively, $\boldsymbol{E}(t)$ is the temporal profile of the MIR electric field, $\varepsilon_g(\boldsymbol{k})$ and $\boldsymbol{D}_{cv}(\boldsymbol{k})$ are the bandgap energy and the transition dipole moment depending on crystal momentum $\boldsymbol{k} = (k_x, k_y)$ in two-dimensional Brillouin zone. Here, we consider the situation in which the transition dipole moment is constant ($\boldsymbol{D}_{cv}(\boldsymbol{k}) = D_0$ and the bandgap energy has a cosine dispersion of the form $\varepsilon_g(\boldsymbol{k}) = \delta - \varepsilon_t \cos(k_x a) \cos(k_y a)$ as shown in Fig. S8 (a). We define the current operator $\hat{\boldsymbol{j}}(t)$ as

$$\hat{\boldsymbol{j}}(t) = -\frac{e\hat{\boldsymbol{p}}_{kin}}{m} = ie\frac{[\hat{\boldsymbol{r}}, \hat{H}(t)]}{\hbar}. \quad (S6)$$

Then, the total current induced by the MIR electric field can be written by

$$\boldsymbol{J}(t) = Tr[\hat{\rho}(t)\hat{\boldsymbol{j}}(t)] = \sum_{\boldsymbol{k}_0} Tr[\tilde{\rho}(\boldsymbol{k}_0, t)\tilde{\boldsymbol{j}}(\boldsymbol{k}_0, t)]. \quad (S7)$$

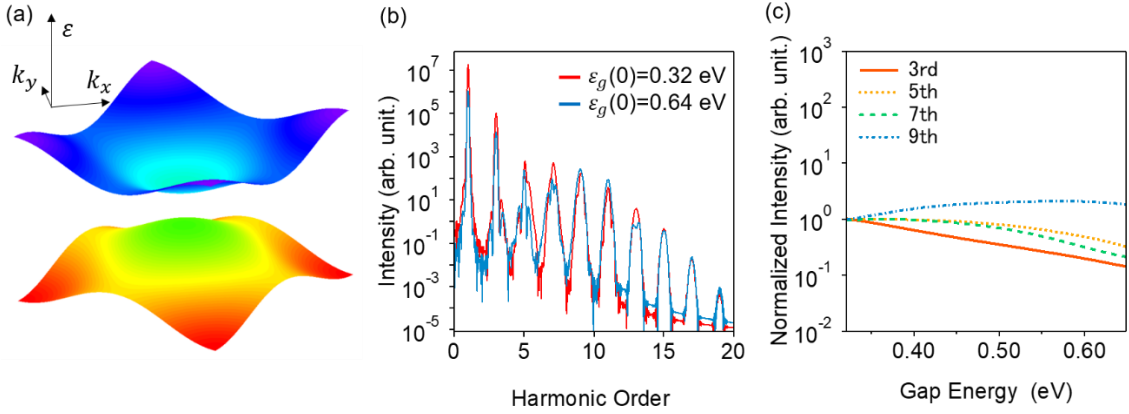

Fig. S8 (a) Schematic of the band structure of two-dimensional two bands with cosine dispersion. (b) The calculated HHG spectra with $\varepsilon_g(0) = 0.32$ eV (red line, $\delta = 1.57$ eV) and $0.64$ eV (blue line, $\delta = 1.89$ eV). (c) Normalized yields of the third (orange solid line), fifth (yellow dotted line), seventh (green dashed line), and ninth (blue dashed-dotted line) as a function of gap energy $\varepsilon_g(0)$. HHG yields are normalized by the value with $\varepsilon_g(0) = 0.32$ eV.

Here, the summation of $\mathbf{k_0}$ runs over the two-dimensional Brillouin zone. The HHG spectrum is given by the magnitude square of the Fourier transformation of the total current $J(t)$.

We use a vector potential with Gaussian envelope

$$A(t) = A_0 \exp\left(-\frac{t^2}{\tau^2}\right) \cos(\Omega_{\text{MIR}} t). \tag{S8}$$

Here, we set $\tau = 95$ fs, $\hbar\Omega_{\text{MIR}} = 0.26$ eV, $A_0 = 0.5\pi\hbar/ea$, $D_0 A_0 \Omega_{\text{MIR}} = 0.08$ eV, $\varepsilon_t = 1.25$ eV, and $\gamma = 0.07$ eV. We calculated the HHG signal by varying $\delta$, which only changes the offset of the gap energy.

Figure S8 (b) shows typical HHG spectra with $\varepsilon_g(0) = 0.32$ eV and $0.64$ eV, which correspond to the optical gap energies of $Ca_2RuO_4$ at 290 K and 50 K, respectively. The calculated HHG spectra, which describes electron dynamics in semiconductors, could not reproduce the experimental data of $Ca_2RuO_4$ shown in Fig. 1(c) in the main text. In particular, we do not observe any no enhancement of higher-order harmonics upon increasing gap energy.

Figure S8 (c) shows the HHG intensities as a function of band gap energy $\varepsilon_g(0)$. Here, each harmonic intensity is normalized by its maximum value. The third harmonic monotonically decreases with an increment of gap energy. In contrast, harmonics greater than the fifth-order show non-monotonic behavior as a function of gap energy.

This result indicates that simple gap opening in the band structure does not enhance higher-order nonlinear optical process as observed in $Ca_2RuO_4$.

We also check the effect of bandwidth (hopping energy $\varepsilon_t$) on HHG yields. Here, we set $\delta = 1.1$ eV and vary $\varepsilon_t$, which also results in bandgap energy shift.

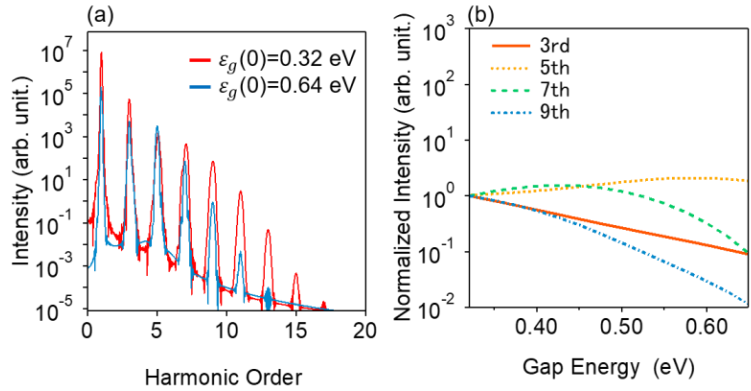

Fig. S9 (a) The calculated HHG spectra with $\varepsilon_g(0) = 0.32$ eV (red line, $\varepsilon_t = 0.78$ eV) and $0.64$ eV (blue line, $\varepsilon_t = 0.46$ eV). (b) Normalized yields of the third (orange solid line), fifth (yellow dotted line), seventh (green dashed line), and ninth (blue dashed-dotted line) as a function of gap energy $\varepsilon_g(0)$. HHG yields are normalized by the values with $\varepsilon_g(0) = 0.32$ eV.

Figure S9 (a) shows typical HHG spectra with $\varepsilon_g(0) = 0.32$ eV and $0.64$ eV, which correspond to the optical gap energy of $Ca_2RuO_4$ at 290 K and 50 K, respectively. Almost all harmonic yields show suppression with a decrease of bandwidth (an increase of bandgap energy). Figures S9 (b) shows the HHG intensities as a function of band gap energy $\varepsilon_g(0)$. These results are qualitatively similar to those with the change of bandgap energy offset: the third harmonic monotonically decreases, whereas fifth and higher harmonics show non-monotonic behavior depending on bandwidth. The amplitude of nonlinear intraband current is proportional to bandwidth for a cosine band structure. Also, kinetic energy of electron-hole pairs under laser field decreases with a reduction of bandwidth. Therefore, bandgap change associated with a reduction of bandwidth also cannot explain our experimental results.

## 7. Bandgap energy dependence of HHG intensities in InAs

Figure S10 shows the HHG intensities as a function of bandgap energy in InAs.

To plot this curve, we used the temperature dependence of InAs bandgap energy from literature [8]. The changes of the harmonic yields depending on bandgap energy are negligible in the observed region. This result is opposite to that in $Ca_2RuO_4$, which shows strong enhancement in HHG yields.

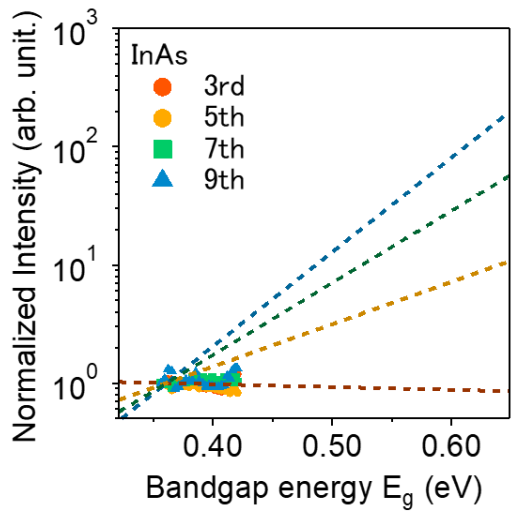

Fig. S10 HHG intensities as a function of bandgap energy in InAs [8]. Dashed lines show the fitting results of HHG yields in $Ca_2RuO_4$.